# A Generalized Kinetic Model for Heterogeneous Gas-Solid Reactions

Zhijie Xu[1,a], Xin Sun[1] and Mohammad A Khaleel[1]

1. Computational Sciences and Mathematics Division, Fundamental and Computational Sciences Directorate, Pacific Northwest National Laboratory, Richland, WA 99352, USA

**Abstract**

We present a generalized kinetic model for gas-solid heterogeneous reactions taking place at the interface between two phases. The model studies the reaction kinetics by taking into account the reactions at the interface, as well as the transport process within the product layer. The standard unreacted shrinking core model relies on the assumption of quasi-static diffusion that results in a steady-state concentration profile of gas reactant in the product layer. By relaxing this assumption and resolving the entire problem, general solutions can be obtained for reaction kinetics, including the reaction front velocity and the conversion (volume fraction of reacted solid). The unreacted shrinking core model is shown to be accurate and in agreement with the generalized model for slow reaction (or fast diffusion), low concentration of gas reactant, and small solid size. Otherwise, a generalized kinetic model should be used.

[a] Electronic mail: zhijie.xu@pnnl.gov Tel: 509-372-4885

## I. Introduction

There are tremendous applications of gas-solid reactions in metallurgical and chemical industries, where reactants are composed of gas and solid phases that undergo chemical changes at their interfaces. Several reaction models in standard textbooks[1] can be used to model typical heterogeneous gas-solid reactions. Particularly, the unreacted shrinking core model that is commonly used for many gas-solid reaction systems has significantly improved our understanding of reaction kinetics. In the standard unreacted shrinking core model, the reaction product forms a solid layer that allows diffusion of gas reactant toward the interface between the product layer and unreacted core. As a result, continuous reaction leads to the advancing of a reaction front or, equivalently, a shrinking core. The entire reaction kinetics involves three steps, showing that reaction proceeds through the continuous inward diffusion of gas reactant and reactions at the interface as shown in Fig. 1. These three steps are:

1. Transport and dissolution of the gas reactant at the solid surface;
2. Diffusion of the gas reactant through the product layer toward the unreacted core;
3. Chemical reaction of the gas with the solid reactant at the product-core interface.

Steps 1, 2, and 3 are demonstrated in Fig. 1, where continuous transport and reaction lead to a moving reaction front at a velocity of $V_s$. The physical description of general gas-solid reactions is fundamentally similar to the "oxidation" process, depicted in classical models generated in the early 1920s[2] by Tammann and Pilling and Bedworth.[3] They established the parabolic oxidation rate law that modern oxidation theory is based upon.[4,5,6] Classical oxidation theory considers the diffusion of a chemical species (mainly oxygen) through the oxide layer as the rate-limiting process. It is well known that for thick oxide films, the rate of oxide growth follows parabolic law, a characteristic of diffusion-limited process where the oxidant must travel an increasingly longer distance to reach the oxide-material interface with increasing oxide thickness. The kinetics of oxide formation under the diffusion-controlled condition was described by Rhines,[7] Darken,[8] and Wagner[9,10] that leads to a parabolic growth law.

In contrast to oxidation, a gas-solid reaction might not be a diffusion-limited process, depending on the competition between transport of the gas reactant and the reaction at the interface. The unreacted shrinking core model is based on the assumption of quasi-static diffusion approximation, where diffusion (or reaction) is assumed to be fast (or slow) enough to allow a steady-state concentration profile of gas reactant established within the product layer. Although it is not able to represent all the mechanisms associated with gas-solid reactions precisely, this simple model is generally accepted for most practical applications and remains a useful tool for understanding important features and for gaining essential knowledge of the complex gas-solid reactions.

Similar assumptions were also made in the original Deal-Grove model[11] for thermal oxidation of silicon. Widely accepted since 1965, the Deal-Grove oxidation model has shown to be accurate over a range of temperatures, oxide thicknesses, and oxidant partial pressures. Despite its success, the Deal-Grove model's validity has been a subject of discussions, which led to the model's re-examination by relaxing the quasi-static diffusion assumption in the original Deal-Grove model.[12]

This paper presents a generalized model and solutions for understanding the gas-solid reaction by accounting for both the transport processes taking place in the product layer and the chemical reactions at the interface. Such transport processes and reactions dominate kinetics gas-solid reactions, i.e., the kinetic process that is mathematically similar to the "solute precipitation"[13,14] and "corrosion" processes.[12] By relaxing the quasi-static diffusion approximation used in the standard unreacted shrinking core model and considering all possible mechanisms, we can present the generalized reaction kinetics for gas-solid reactions without any *a priori* assumptions using the reduce-boundary-function method.[12,15] It is shown that the generalized results can be reduced to the unreacted shrinking core model for slow reaction (or fast diffusion), small solid size, and low gas reactant concentration.

## II. Mathematical Model for General Gas-Solid Reactions

The mathematical model provides the governing equations for general gas-solid reactions. This model includes the diffusion of gas reactant in the solid layer and reactions at the product-layer/unreacted-core interface ($\Gamma_2$ in Figure 1). The dynamics of the moving reaction front is a result of the competition between the transport of gas reactant to the interface and the consumption of gas reactant due to the chemical reaction at the interface. Consider a general gas-solid reaction of type:

$$A(g) + bB(s) \xrightarrow{k} cC(s), \tag{1}$$

where $C$ is the reaction product accumulated on the unreacted core and $b$ and $c$ are the stoichiometric coefficients. Equations

$$\partial C_A / \partial t = D_A \nabla^2 C_A \tag{2}$$

in a Cartesian coordinate or

$$\frac{\partial C_A}{\partial t} = D_A \left( \frac{\partial^2 C_A}{\partial r^2} + \frac{d}{r} \frac{\partial C_A}{\partial r} \right) \tag{3}$$

in spherical coordinate describe the diffusive transport of gas reactant $A$ within the product layer (gray color in Figure 1), where $C_A(r,t)$ is the concentration of gas species $A$ at any given position $r$ and time $t$, $\nabla$ is the Laplace operator in Eq. (2), and $D_A$ is the diffusion coefficient of gas reactant in the product layer (gray color in Figure 1). Equation (3) represents the diffusion for different geometries with $d$=0 corresponding to a one-dimensional diffusion, $d$=1 corresponding to the diffusion in a cylindrical geometry, and $d$=2 corresponding to the diffusion in a spherical geometry. In Figure 1, $R$ is the initial radius of the unreacted solid, and $R_c$ is the radius of the unreacted shrinking core.

The flux of gas reactant into interface $\Gamma_2$ in Figure 1 should balance the gas consumption due to the gas-solid reaction. By assuming the gas-solid reaction at the interface in the form of Eq. (1), the simplest kinetic rate equation that we can use is:

$$\frac{d\left(C_c\right)}{dt} = ck_1\left(C_A\big|_2^+\right)\left(C_B\big|_2^+\right)^b \tag{4}$$

based on the law of mass action, where $k_1$ is the reaction rate constant. The left side of Eq. (4) shows the generation rate of the reaction product $C$, where $C_c$ is the product concentration on $\Gamma_2$ with a unit of $mol/m^2$. On the right-hand side, $C_{(\square)}\big|_2^+$ is the interface concentration on the positive side of interface $\Gamma_2$ with "$\square$" representing either $A$ or $B$. Therefore, the moving velocity of reaction front is given by:

$$V_s = -\frac{dR_c}{dt} = \frac{cD_A}{\rho}\frac{\partial C_A}{\partial r}\bigg|_2^+ = \frac{ck_1\left(C_B\big|_2^+\right)^b}{\rho}\left(C_A\big|_2^+\right) \text{ at } \Gamma_2 \ (r = R_c) \tag{5}$$

as a result of the local mass conservation, where $\rho$ is the molar density of the reaction product $C$, and $R_c$ is the radius of unreacted solid core (black color in Figure 1). The flux of the gas reactive species from the gas phase (white color in Figure 1) to the outer surface $\Gamma_1$ is expressed as:

$$H_0\left(C^* - C_A\big|_1^-\right) = D_A \frac{\partial C_A}{\partial r}\bigg|_1^- \text{ at } \Gamma_1 \ (r = R), \tag{6}$$

and it equals the flux of the diffusion within the product layer (gray color in Figure 1), where $H_0$ is a gas-phase transport coefficient, and $C^*$ is the equilibrium concentration of the gas reactant $A$ in the product layer (or the gas reactant solubility in the solid phase). $C_A\big|_1^-$ is the concentration of $A$ on the negative side of the interface $\Gamma_1$. This relationship is an analogous to the Newton's law of cooling. The equilibrium concentration $C^*$ is related to the partial pressure $P_{gas}$ of the gas reactant in the gas phase at the external surface of solid phase (positive side of interface $\Gamma_1$) through the constant $k_2$ from Henry's law,

$$C^* = k_2 P_{gas}. \tag{7}$$

The preceding system of equations (Eq. (3) with the interface conditions Eqs. (5)-(7)) can be rewritten in dimensionless form by introducing the unit of length $R$ (the characteristic size of the solid); unit of time $S = R^2/D_A$; unit of velocity $U = D_A/R$; dimensionless numbers $D_a = k_1\left(C_B\big|_2^+\right)^b R\big/D_A$, representing the ratio between reaction and diffusion; and $h_0 = H_0 R/D_A$, representing the ratio between gas-phase transport and diffusion. The new system of equations reads:

$$\frac{\partial c_A}{\partial t} = \frac{\partial^2 c_A}{\partial r^2} + \frac{d}{r}\frac{\partial c_A}{\partial r}, \tag{8}$$

$$\frac{\partial c_A}{\partial r}\bigg|_1^- = h_0\left(c^* - c_A\big|_1^-\right) \text{ at the } \Gamma_1 \ (r = 1), \text{ and} \tag{9}$$

$$v_s = -\frac{dr_c}{dt} = \frac{dx_c}{dt} = c\frac{\partial c_A}{\partial r}\bigg|_2^+ = cD_a\left(c_A\big|_2^+\right) \text{ at the } \Gamma_2 \ (r = r_c), \tag{10}$$

where $c_A = C_A/\rho$ is the normalized concentration by $\rho$, the molar density of reaction product *C*. $r_c = R_c/R$ is the normalized size of unreacted core, and $x_c = 1 - r_c$ is the normalized distance of the proceeding reaction front. Solutions from Eqs. (8)-(10) are only dependent on the dimensionless numbers $D_a$ and $h_0$, geometry constant *d*, and the normalized equilibrium concentration $c^* = k_2 P_{gas}/\rho$. Dimensionless number $D_a$ represents the ratio between reaction and diffusion with $D_a \to \infty$ corresponding to the diffusion-limited regime and $D_a \to 0$ corresponding to the reaction-limited regime. The effects of concentration of solid reactant *B* and characteristic size *R* (solid size) also were grouped into $D_a$. In principle, $D_a$ is a function of temperature because both reaction rate $k_1$ and diffusion coefficient $D_A$ are temperature dependent.

## III. Solutions with Steady-State Diffusion Approximation

Initially, we consider solutions available based on the steady-state diffusion approximation, namely the unreacted shrinking core model where time derivative of concentration vanishes in Eqs. (3) and (8). In this case, the original Eq. (8) for spherical particles (*d* =2) is reduced to:

$$\frac{\partial^2 c_A}{\partial r^2} + \frac{2}{r}\frac{\partial c_A}{\partial r} = 0, \qquad (11)$$

with boundary conditions:

$$c_A = c_A\big|_1^- \text{ at } \Gamma_1, \text{ where } r = 1, \qquad (12)$$

and $c_A = c_A\big|_2^+$ at $\Gamma_2$, where $r = r_c$. (13)

The solution to Eq. (11) with boundary conditions (12) and (13) is readily determined as:

$$c_A = -\frac{1}{r}\left(c_A\big|_1^- - c_A\big|_2^+\right)\Big/\left(\frac{1}{r_c}-1\right) + \left(c_A\big|_1^- - r_c\, c_A\big|_2^+\right)\Big/\left(1-r_c\right). \qquad (14)$$

At interface $\Gamma_1$, the flux

$$F_1 = 4\pi h_0\left(c^* - c_A\big|_1^-\right) = F_2 = 4\pi\frac{\partial c_A}{\partial r}\bigg|_1^- \qquad (15)$$

can be obtained from Eq. (9). Due to the steady-state diffusion approximation, we have flux

$$F_2 = 4\pi\frac{\partial c_A}{\partial r}\bigg|_1^- = F_3 = 4\pi r_c^2\frac{\partial c_A}{\partial r}\bigg|_2^+ = 4\pi\left(c_A\big|_1^- - c_A\big|_2^+\right)\Big/\left(\frac{1}{r_c}-1\right) \qquad (16)$$

by solving Eq. (11). At interface $\Gamma_2$, the flux due to the gas-solid reaction is:

$$F_4 = 4\pi r_c^2 D_a\, c_A\big|_2^+ = 4\pi r_c^2 v_s/c\,. \qquad (17)$$

All four fluxes should be equal to each other due to the steady-state diffusion approximation, and are assumed to be *F*. From Eqs. (10) and (17), we have

$$F = F_1 = F_2 = F_3 = F_4 = 4\pi r_c^2 v_s/c\,, \qquad (18)$$

and the moving velocity of reaction front is

$$v_s = \frac{cc^*}{\frac{r_c^2}{h_0}+\frac{1}{D_a}+r_c\left(1-r_c\right)}\,. \qquad (19)$$

After introducing the conversion defined as $\xi = 1 - r_c^3$, the integration of Eq. (19) leads to the following kinetic expressions between reaction time *t* and $r_c$ or $\xi$:

$$t = \frac{1}{cc^*}\left[\frac{1}{3h_0}\left(1-r_c^3\right)+\frac{1}{D_a}\left(1-r_c\right)+\frac{1}{6}\left(1-3r_c^2+2r_c^3\right)\right], \qquad (20)$$

and

$$t = \frac{1}{cc^*}\left[\frac{1}{3h_0}\xi + \frac{1}{D_a}\left(1-(1-\xi)^{1/3}\right) + \frac{1}{6}\left(1-3(1-\xi)^{2/3} + 2(1-\xi)\right)\right]. \tag{21}$$

The total time needed for a complete conversion is:

$$t_c = t(r_c = 0) = \frac{1}{cc^*}\left(\frac{1}{3h_0} + \frac{1}{D_a} + \frac{1}{6}\right). \tag{22}$$

Similar results can be obtained for solids with cylindrical geometry ($d$=1), where the concentration profile is:

$$c_A = -\frac{1}{r}\left(c_A\big|_1^- - c_A\big|_2^+\right)\Big/\left(\frac{1}{r_c} - 1\right) + \left(c_A\big|_1^- - r_c\, c_A\big|_2^+\right)\Big/(1-r_c). \tag{23}$$

The moving velocity of the reaction front is:

$$v_s = -\frac{dr_c}{dt} = \frac{cc^*}{\dfrac{r_c}{h_0} + \dfrac{1}{D_a} - r_c \log(r_c)}. \tag{24}$$

The kinetic expressions for the variation of $r_c$ or conversion $\xi$ with time $t$ are:

$$t = \frac{1}{cc^*}\left[\frac{1}{h_0}\left(\frac{1-r_c^2}{2}\right) + \frac{1}{D_a}(1-r_c) + \frac{1}{4}(1-r_c^2) + \frac{1}{2}r_c^2 \log(r_c)\right], \tag{25}$$

and

$$t = \frac{1}{cc^*}\left[\frac{1}{h_0}\left(\frac{\xi}{2}\right) + \frac{1}{D_a}\left(1-(1-\xi)^{1/2}\right) + \frac{\xi}{4} + \frac{1-\xi}{2}\log(1-\xi)^{1/2}\right]. \tag{26}$$

The time required for a complete conversion is:

$$t_c = t(r_c = 0) = \frac{1}{cc^*}\left(\frac{1}{2h_0} + \frac{1}{D_a} + \frac{1}{4}\right). \tag{27}$$

We also present the results for one-dimensional planar geometry, where $d$=0. The reaction front velocity is:

$$v_s = -\frac{dr_c}{dt} = \frac{cc^*}{\frac{1}{h_0} + \frac{1}{D_a} + 1 - r_c}. \tag{28}$$

The corresponding kinetic expressions and time for a complete conversion are:

$$t = \frac{1}{cc^*}\left[\frac{1-r_c}{h_0} + \frac{1}{D_a}\left(1-r_c\right) + \frac{1}{2}\left(1+r_c^2\right) - r_c\right], \tag{29}$$

$$t = \frac{1}{cc^*}\left[\frac{\xi}{h_0} + \frac{\xi}{D_a} + \frac{1+\left(1-\xi\right)^2}{2} - 1 + \xi\right], \tag{30}$$

and

$$t_c = t\left(r_c = 0\right) = \frac{1}{cc^*}\left(\frac{1}{h_0} + \frac{1}{D_a} + \frac{1}{2}\right). \tag{31}$$

For all three cases with different *d*, it is observed that expressions for the reaction front velocity $v_s$ and complete time $t_c$ are generally similar. Figure 3 presents the variation of $t_c$ with $D_a$ for all three geometries. $v_s$ (or $t_c$) is directly (inversely) proportional to the normalized equilibrium concentration $c^*$, which is an important conclusion from the unreacted shrinking core model. Intuitively, this proportionality, true for small $c^*$, would not apply at large $c^*$ or $D_a$ as we continuously increase $c^*$ because the steady-state diffusion approximation is not valid at large $c^*$ and $D_a$. This is confirmed by our generalized model in the following section.

## IV. General Solutions for the Gas-solid Reaction

In this section, we will solve Eqs. (8)-(10) by relaxing the steady-state diffusion approximation. As depicted in Figure 2, we first introduce the following relationships between the interface values and interface velocity through a straightforward differential analysis:

$$\left.\frac{\partial c_A}{\partial t}\right|_2^+ = \frac{\partial \left.c_A\right|_2^+}{\partial t} + \left.\frac{\partial c_A}{\partial r}\right|_2^+ \cdot v_s \text{ and} \tag{32}$$

$$\left.\frac{\partial\left(\partial c_A/\partial r\right)}{\partial t}\right|_2^+ = \frac{\partial\left(\partial c_A/\partial r\right)\Big|_2^+}{\partial t} + \left.\frac{\partial^2 c_A}{\partial r^2}\right|_2^+ \cdot v_s\,. \tag{33}$$

Similarly, higher order derivatives (*n*th order) at the interface can be obtained in the same fashion:

$$\left.\frac{\partial\left(\partial^n c_A/\partial r^n\right)}{\partial t}\right|_2^+ = \frac{\partial\left(\partial^n c_A/\partial r^n\right)\Big|_2^+}{\partial t} + \left.\frac{\partial^{n+1} c_A}{\partial r^{n+1}}\right|_2^+ \cdot v_s\,. \tag{34}$$

From the diffusion Eq. (8), we obtain the following expression:

$$\frac{\partial^{n+1} c_A}{\partial t \partial r^n} = \frac{\partial^{n+2} c_A}{\partial r^{n+2}} + \sum_{k=1}^{n+1} \frac{n!(-1)^{n+1-k}}{(k-1)!} \cdot \frac{d}{r^{n+2-k}} \frac{\partial^k c_A}{\partial r^k}\,. \tag{35}$$

Based on Eq. (35), the following relationship between interface values can be obtained:

$$\left.\frac{\partial^{n+2} c_A}{\partial r^{n+2}}\right|_2^+ = \left.\frac{\partial\left(\partial^n c_A/\partial r^n\right)}{\partial t}\right|_2^+ - \sum_{k=1}^{n+1} \frac{n!(-1)^{n+1-k}}{(k-1)!} \cdot \frac{d}{r_c^{\,n+2-k}} \left.\frac{\partial^k c_A}{\partial r^k}\right|_2^+,\; n = 0,1,2,3...... \tag{36}$$

By using Eq. (34), equation (36) can be rewritten as:

$$\left.\frac{\partial^{n+2} c_A}{\partial r^{n+2}}\right|_2^+ = \frac{\partial\left(\partial^n c_A/\partial r^n\right)\Big|_2^+}{\partial t} + \left.\frac{\partial^{n+1} c_A}{\partial r^{n+1}}\right|_2^+ \cdot \left(v_s - \frac{d}{r_c}\right) - \sum_{k=1}^{n} \frac{n!(-1)^{n+1-k}}{(k-1)!} \cdot \frac{d}{r_c^{\,n+2-k}} \left.\frac{\partial^k c_A}{\partial r^k}\right|_2^+,$$

$$n = 0,1,2,3...... \tag{37}$$

By using interface conditions (Eq. (10)), we obtained the equations for interfacial concentration and corresponding derivatives up to the third order:

$$c_A\big|_2^+ = v_s/(cD_a), \tag{38}$$

$$\left.\frac{\partial c_A}{\partial r}\right|_2^+ = \frac{v_s}{c}, \tag{39}$$

$$\left.\frac{\partial^2 c_A}{\partial r^2}\right|_2^+ = \left.\frac{\partial c_A}{\partial t}\right|_2^+ - \frac{d}{r_c}\left.\frac{\partial c_A}{\partial r}\right|_2^+ = \frac{1}{cD_a}\frac{\partial v_s}{\partial t} + \frac{v_s}{c}\left(v_s - \frac{d}{r_c}\right), \text{ and} \tag{40}$$

$$\left.\frac{\partial^3 c_A}{\partial r^3}\right|_2^+ = \frac{1}{c}\left[\frac{1}{D_a}\left(v_s - \frac{d}{r_c}\right) + 1\right]\frac{\partial v_s}{\partial t} + \frac{v_s}{c}\left(v_s - \frac{d}{r_c}\right)^2 + \frac{v_s}{c}\frac{d}{r_c^2}. \tag{41}$$

In principle, any higher order concentration derivatives ( $\left.\frac{\partial^4 c_A}{\partial r^4}\right|_2^+, \left.\frac{\partial^5 c_A}{\partial r^5}\right|_2^+, \ldots\ldots$ ) can be obtained in a similar manner using Eq. (37). We can express the concentration $c_A$ in terms of those derivatives via Taylor expansion:

$$c_A(x) = c_A\big|_2^+ + \sum_{n=1}^{\infty}\frac{x^n}{n!}\left.\frac{\partial^n c_A}{\partial r^n}\right|_2^+ = c_A\big|_2^+ + x\left.\frac{\partial c_A}{\partial r}\right|_2^+ + \frac{x^2}{2!}\left.\frac{\partial^2 c_A}{\partial r^2}\right|_2^+ + \frac{x^3}{3!}\left.\frac{\partial^3 c_A}{\partial r^3}\right|_2^+ + \ldots\ldots, \tag{42}$$

where $x$ is the distance from the interface $\Gamma_2$. Substitution of Eqs. (38)-(41) into Taylor expansion in Eq. (42) and using relationship

$$\frac{\partial v_s}{\partial t} = \frac{\partial v_s}{\partial x_c}\frac{\partial x_c}{\partial t} = v_s\frac{\partial v_s}{\partial x_c} \tag{43}$$

lead to the concentration of gas reactant within the product layer:

$$c_A(x) \approx \frac{1}{c}\left\{\frac{v_s}{D_a} + \gamma\left[\exp\left(\frac{v_s}{\gamma}x\right) - 1\right] + \frac{\gamma^2}{D_a v_s}\frac{\partial v_s}{\partial x_c}\left[\exp\left(\frac{v_s}{\gamma}x\right) - \frac{v_s}{\gamma}x - 1\right]\right\}, \tag{44}$$

where

$$\gamma = \frac{1}{1 - d/(r_c v_s)} \tag{45}$$

represents the effect of geometry (i.e., different values of $d$). The derivative can be obtained from Eq. (44) for any given $x$:

$$\frac{\partial c_A}{\partial x}=\frac{1}{c}\left\{\left(v_s+\frac{\gamma}{D_a}\frac{\partial v_s}{\partial x_c}\right)\exp\left(\frac{v_s}{\gamma}x\right)-\frac{\gamma}{D_a}\frac{\partial v_s}{\partial x_c}\right\}. \tag{46}$$

Considering the boundary condition (9) that can be rewritten as:

$$c^*=\frac{1}{h_0}\frac{\partial c_A}{\partial r}\bigg|_1^- + c_A\big|_1^- \text{ at } x=x_c, \tag{47}$$

we are able to arrive at the final expression:

$$\frac{\partial v_s}{\partial x_c}=-\frac{\gamma-\exp\left(-v_s x_c/\gamma\right)\left[cc^*-\left(v_s/D_a\right)+\gamma\right]+v_s/h_0}{\gamma^2\left[1-\left(1+v_s x_c/\gamma\right)\exp\left(-v_s x_c/\gamma\right)\right]+\gamma v_s\left[1-\exp\left(-v_s x_c/\gamma\right)\right]/h_0}\cdot D_a v_s. \tag{48}$$

If the gas-phase transport coefficient $H_0$ is much larger than the diffusion rate, the dimensionless number $h_0$ can be removed from Eq. (48), resulting in:

$$\frac{\partial v_s}{\partial x_c}=-\frac{\gamma-\exp\left(-v_s x_c/\gamma\right)\left[cc^*-\left(v_s/D_a\right)+\gamma\right]}{\gamma^2\left[1-\left(1+v_s x_c/\gamma\right)\exp\left(-v_s x_c/\gamma\right)\right]}\cdot D_a v_s. \tag{49}$$

Specifically for $c$=1 and the final expression can be written as:

$$\frac{\partial v_s}{\partial x_c}=-\frac{\gamma-\exp\left(-v_s x_c/\gamma\right)\left(c^*-\left(v_s/D_a\right)+\gamma\right)}{\gamma^2\left[1-\left(1+v_s x_c/\gamma\right)\exp\left(-v_s x_c/\gamma\right)\right]}\cdot D_a v_s. \tag{50}$$

By numerically solving Eq. (50), a first-order ordinary differential equation (ODE), with initial condition $v_s\big|_{t=0}=D_a c^*$ and $x_c\big|_{t=0}=0$, the kinetics for gas-solid reaction for solids with different geometries (planar, cylinder, and sphere) can be resolved for any given $c^*$, $D_a$, and $d$.

## V. Comparison with Solutions Based on Steady-state Diffusion Approximation

The unreacted shrinking core model is based on the steady-state diffusion approximation. Although the shrinking core model is accepted as the best simple model for most gas-solid reaction systems, it might not be accurate under certain circumstances where the steady-state diffusion approximation is not valid. A comparison between the solutions obtained in Section III and solutions by numerically solving Eq. (50) for *c*=1 should provide useful insight.

Figure 4 presents a comparison of the normalized velocity $u_s = v_s / \left( D_a c^* \right)$, varying with $x_c$ (the distance of the reaction front's penetration) between the solution with steady-state diffusion approximation (Eq. (24) is represented by the solid line) and the solution from Eq. (50) for *d*=1 and $D_a = 1$. However, Figure 4 includes different $c^*$ (dotted line for $c^* = 1$, dashed line for $c^* = 10$, and dash-dot line for $c^* = 100$). The unreacted shrinking core model with steady-state diffusion approximation (Eq. (24)) predicts that $u_s$ is independent of $c^*$. For small $c^*$, the solution from the generalized model is in good agreement with the predictions from the unreacted shrinking core model. With increasing concentration $c^*$, the discrepancies between the two models also increase. Obviously, $u_s$ is dependent on $c^*$, and the unreacted shrinking model merely is a good approximation for small $c^*$ but overestimates $u_s$ for large $c^*$.

Figures 5 and 6 show a similar comparison between the generalized model and the unreacted shrinking core model for *d*=1 (cylinder geometry), and $D_a = 10$ and 100, respectively. Again, the two models are in better agreement with each other for small $c^*$. The effect of $D_a$ can be studied by comparing Figures 4, 5, and 6. Large $D_a$ means slow diffusion, which invalidates the steady-state diffusion approximation. Hence, the two models are only in good agreement for small $D_a$. We conclude the widely accepted unreacted shrinking core model is a good prediction only for small $c^*$ and $D_a$ where the steady-state diffusion approximations remain valid. Otherwise, that approximation will lead to errors.

For completeness, the numerical solutions for normalized time $t_m = t_c c^* D_a$ ($t_c$ is the time required for a complete conversion) for a variation with $D_a$ are also presented in Figure 7. Based on the steady-state diffusion approximation, time $t_m$ can be obtained from Eq. (27):

$$t_m = t_c c^* D_a = 1 + \frac{1}{4} D_a \,. \qquad (51)$$

For comparison, it was plotted in Figure 7 as the solid line. Numerical results from the generalized model by solving Eq. (50) also are depicted in Figure 7: $c^*$=1 is the dash line and $c^*$=10 is dash dot line, respectively. A large discrepancy is shown as expected for large concentration $c^*$.

## VI. Conclusion

A mathematical model and solutions for a model gas-solid reaction for solids with different geometries (planar, cylinder, and sphere) are presented for given dimensionless number $D_a$ (lumping the effect of reaction rate, diffusion, and solid characteristic size), and equilibrium concentration $c^*$. In comparison with the generalized model, it was shown that the unreacted shrinking core model based on the steady-state diffusion approximation is only valid for small $c^*$ and $D_a$. Therefore, the generalized model offers a much better descriptive range for the heterogeneous gas-solid reaction process for a diverse range of relevant parameters. Ongoing studies will include comparison of the generalized model with numerical modeling and experimental data.

Figure 1. Schematic plot of a typical gas-solid reaction and the gas reactant concentration profile.

Figure 2. Schematic plot of the moving interface used to derive the differential relationships between interface values and interface velocity (Eqs. (32) and (33)).

Figure 3. Variation of $t_c$ (time required for a complete conversion) with $D_a$ (a dimensionless ratio between reaction and diffusion) for solids with different geometries.

Figure 4. Variation of $u_s$ (normalized reaction front velocity) with $x_c$ (reaction front position) for cylindrical geometry (*d*=1), $D_a = 1$, and $c^* = 1, 10, 100$. The solid line represents the solution with steady-state diffusion approximation.

Figure 5. Variation of $u_s$ (normalized reaction front velocity) with $x_c$ (reaction front position) for cylindrical geometry (*d*=1), $D_a = 10$, and $c^* = 1, 10, 100$. The solid line represents the solution with steady-state diffusion approximation.

Figure 6. Variation of $u_s$ (normalized reaction front velocity) with $x_c$ (reaction front position) for cylindrical geometry (*d*=1), $D_a = 100$, and $c^* = 1, 10, 100$. The solid line represents the solution with steady-state diffusion approximation.

Figure 7. Variation of $t_m$ (normalized time required for a complete conversion) with $D_a$ (a dimensionless ratio between reaction and diffusion) for cylindrical geometry (*d*=1) and $c^* = 1$ and 10. The solid line represents the solution with steady-state diffusion approximation.

Figure 1.

(a)

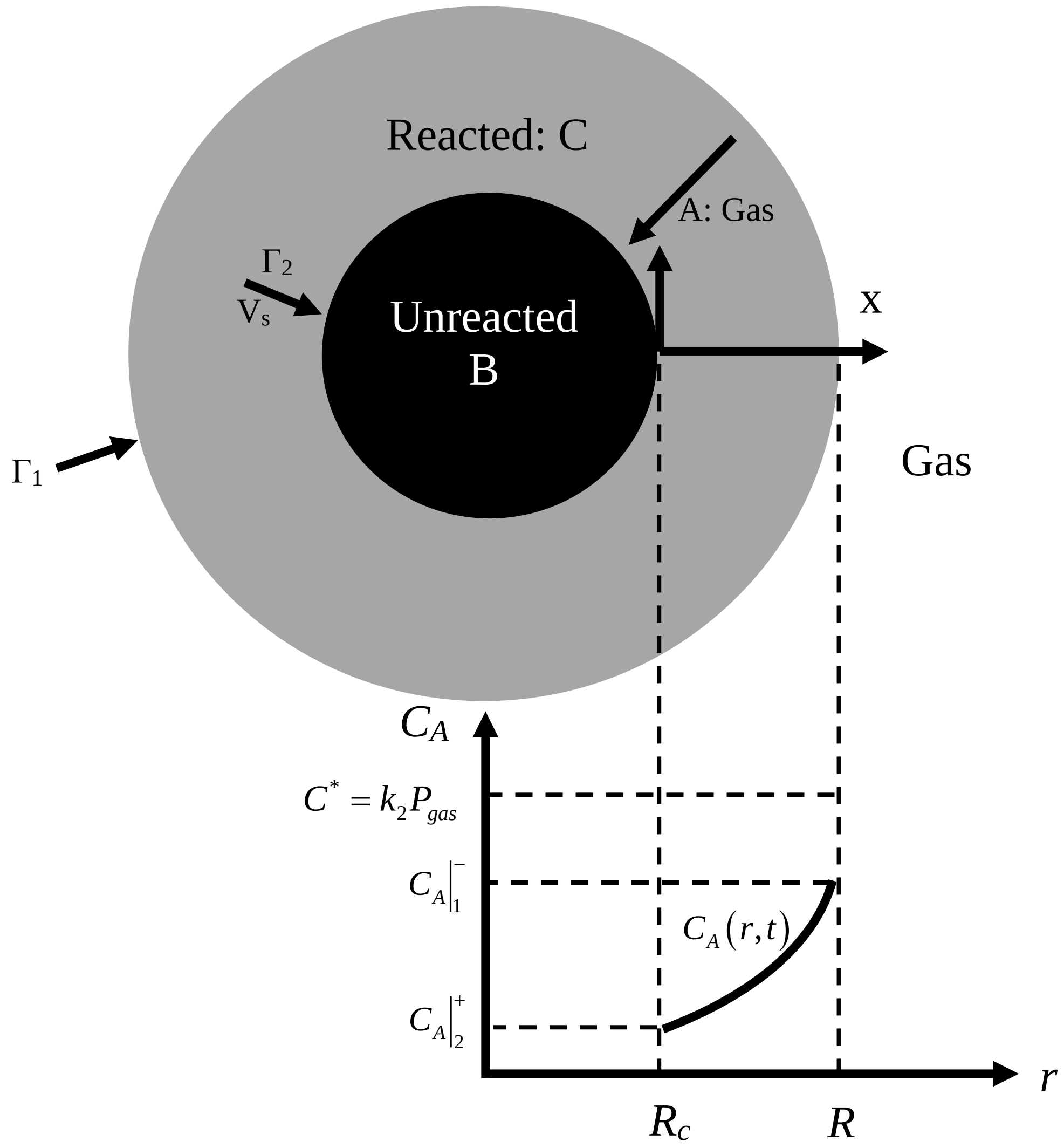

Figure 2.

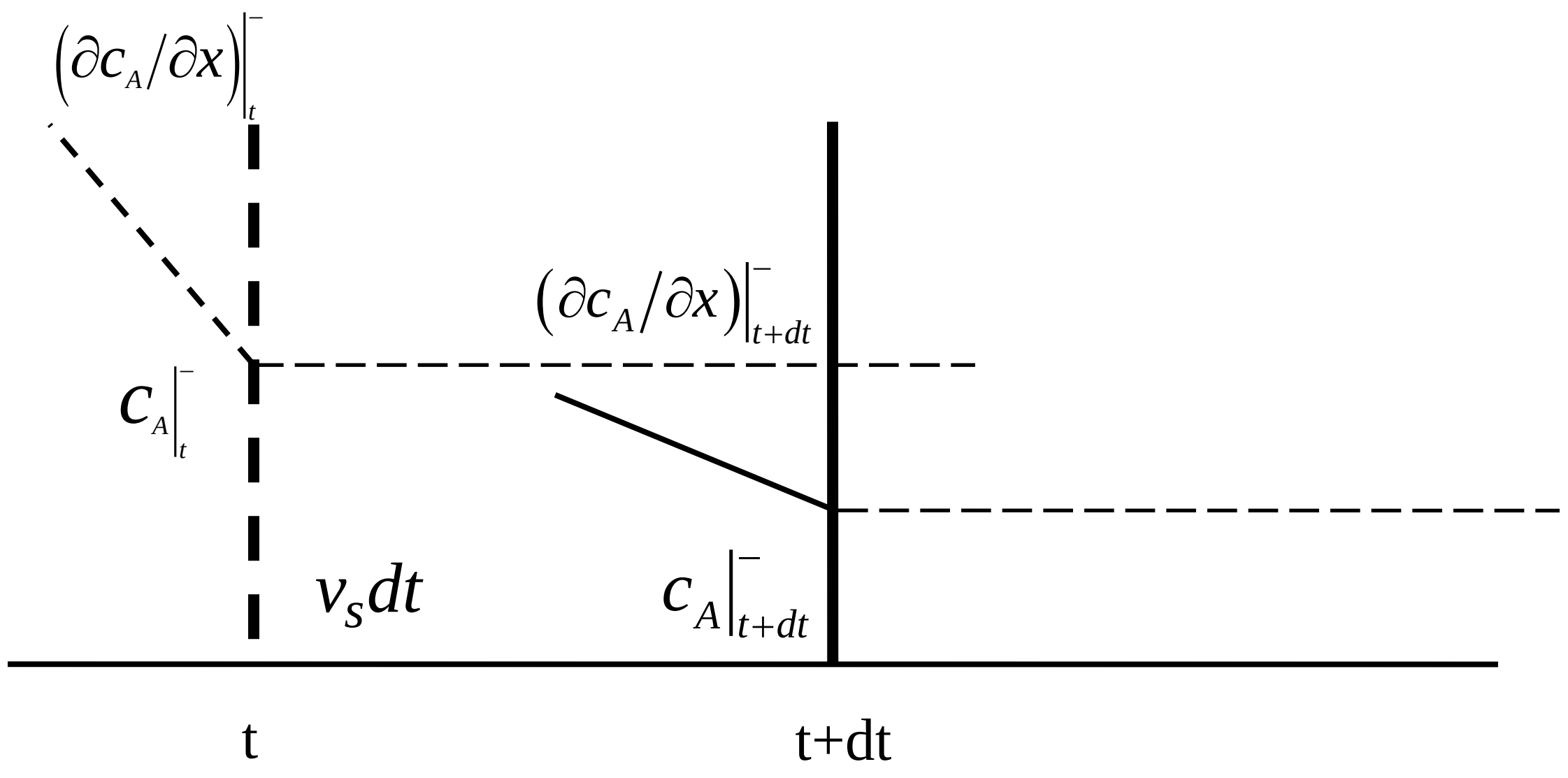

Figure 3.

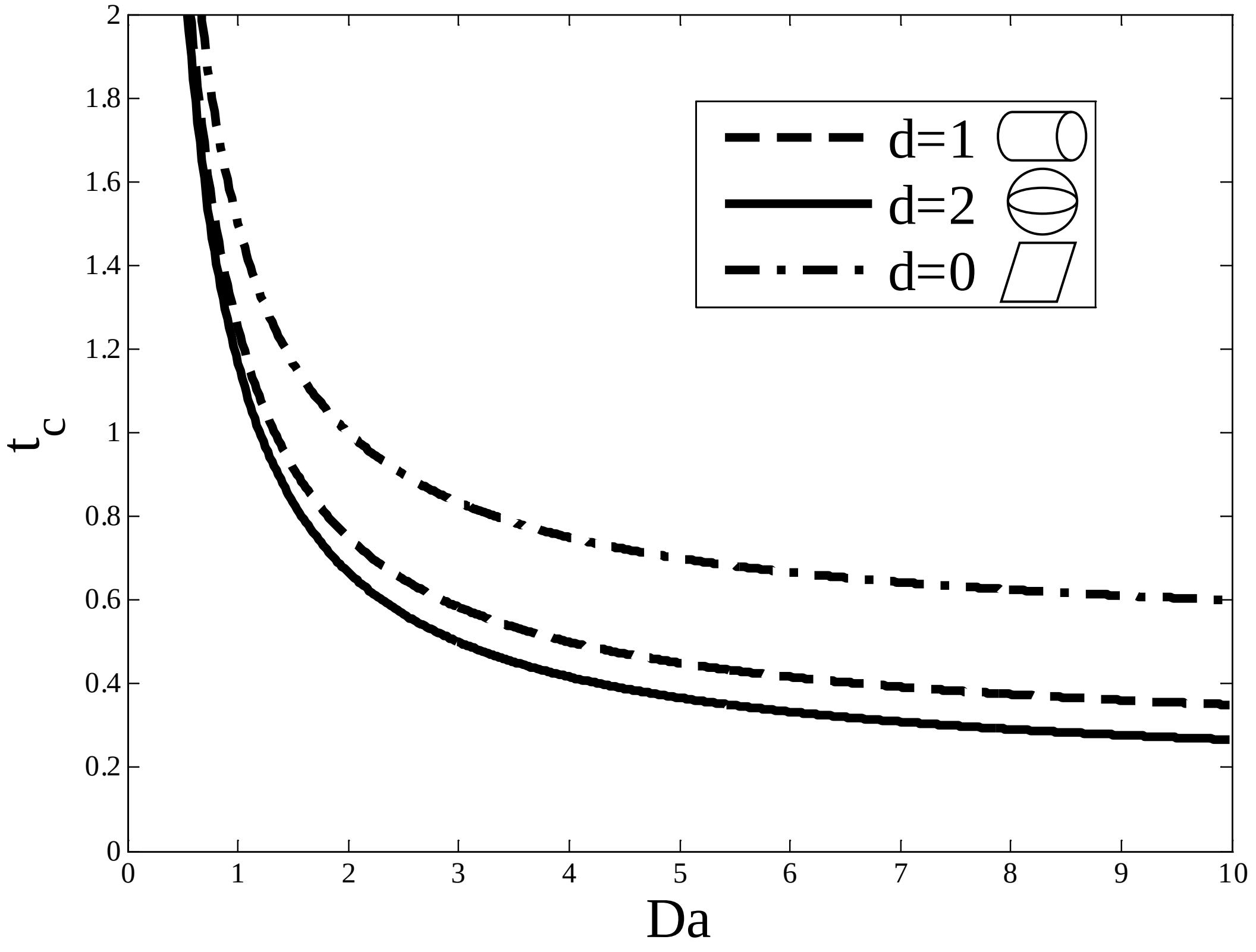

Figure 4.

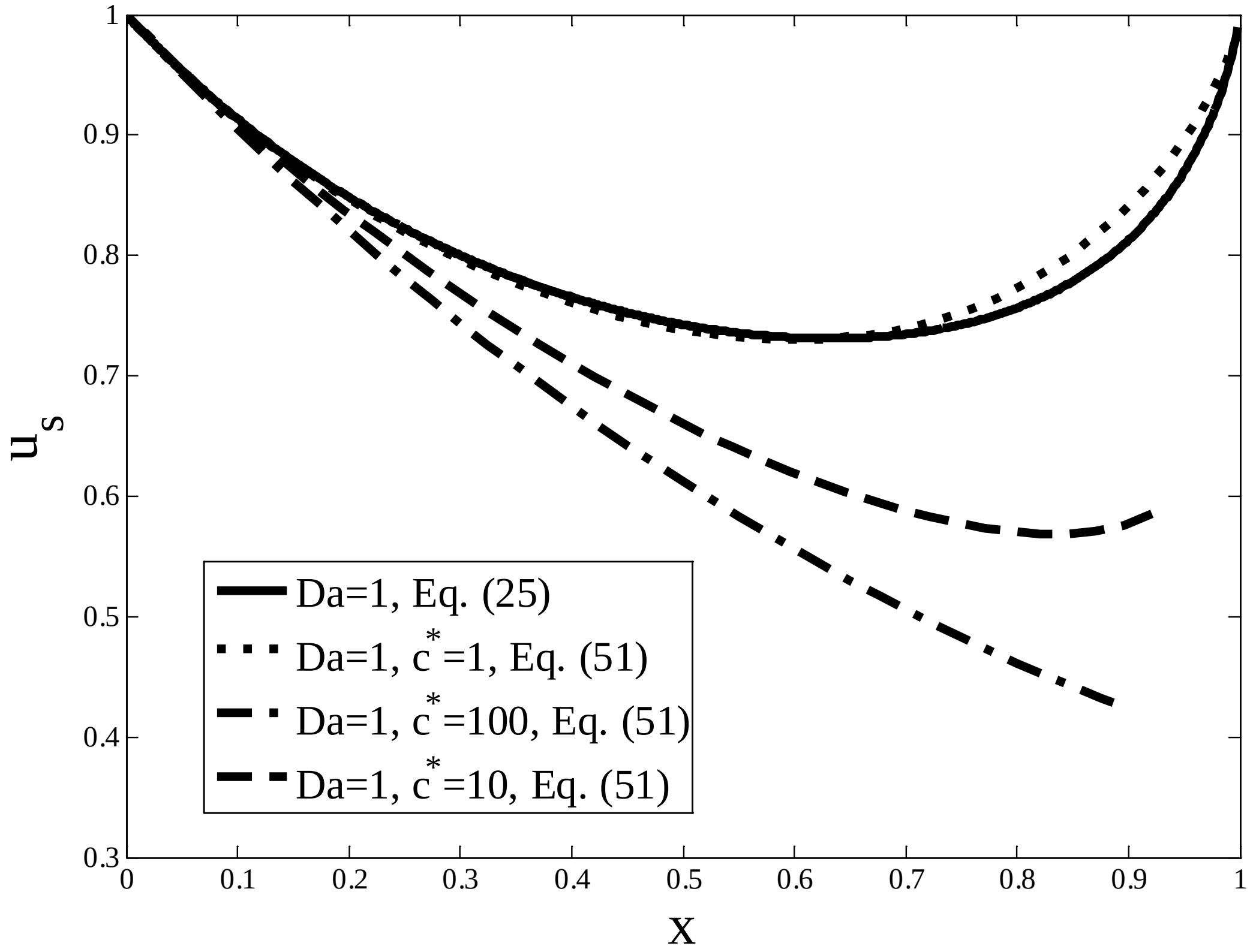

Figure 5.

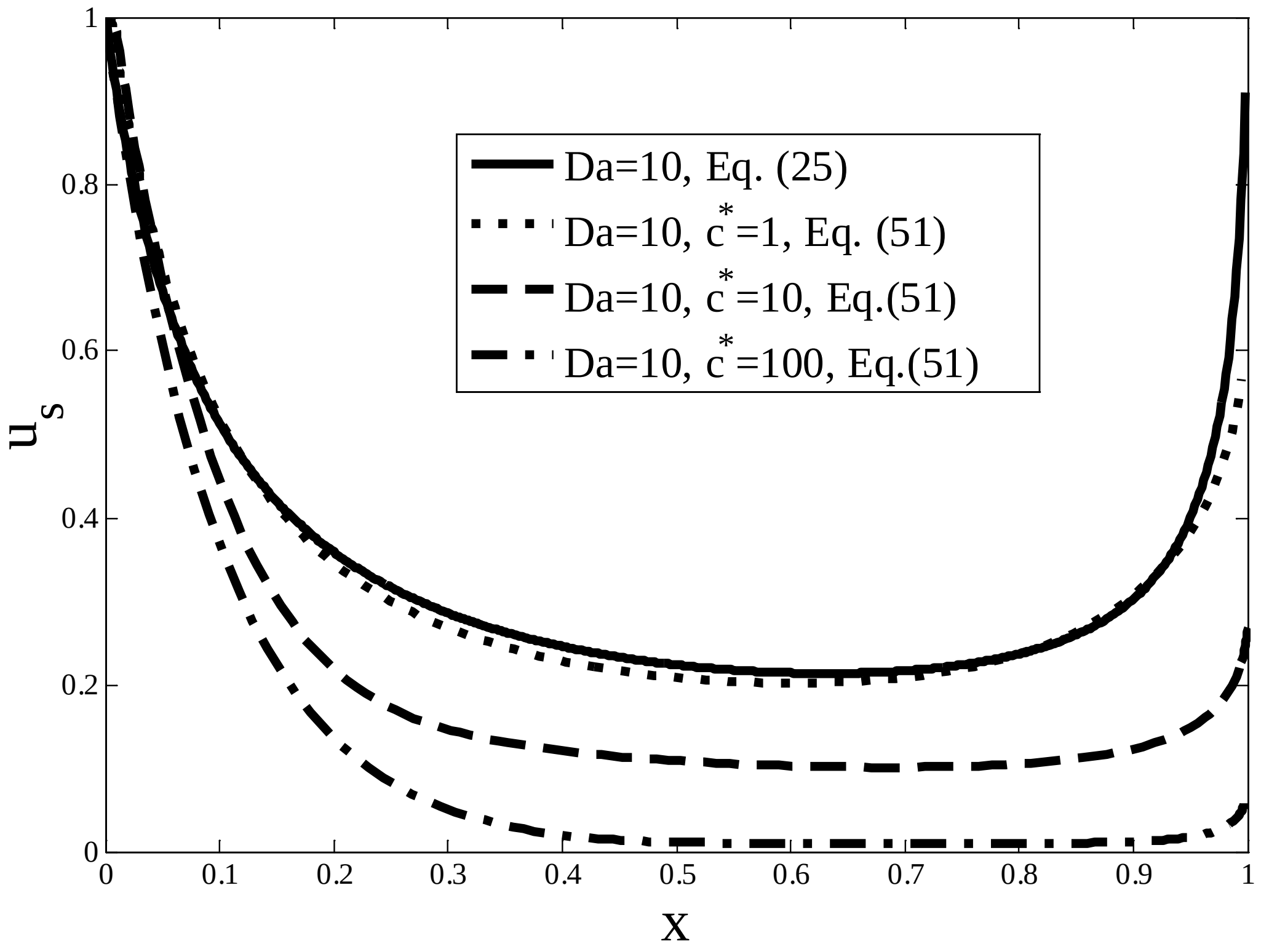

Figure 6

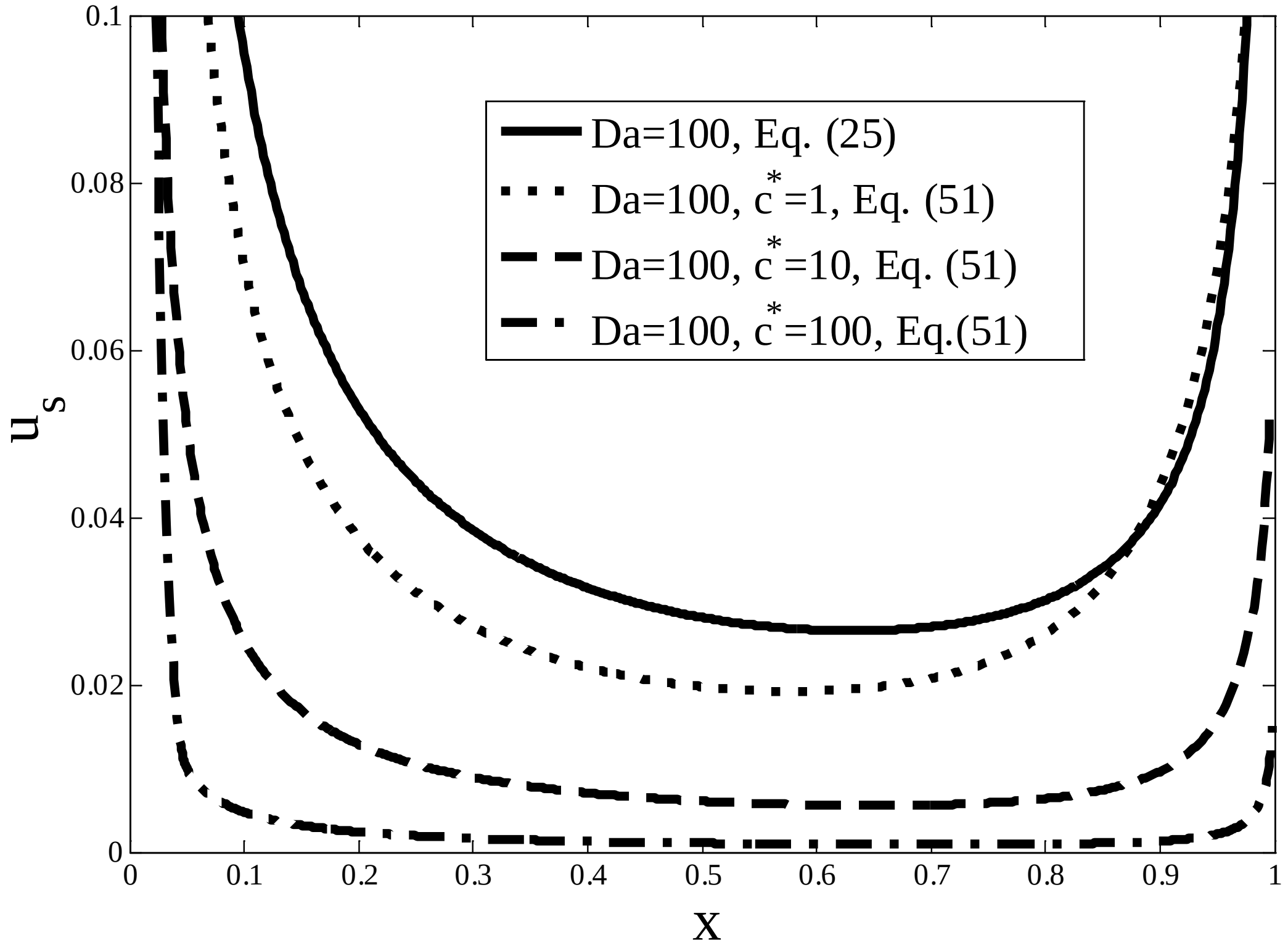

Figure 7.

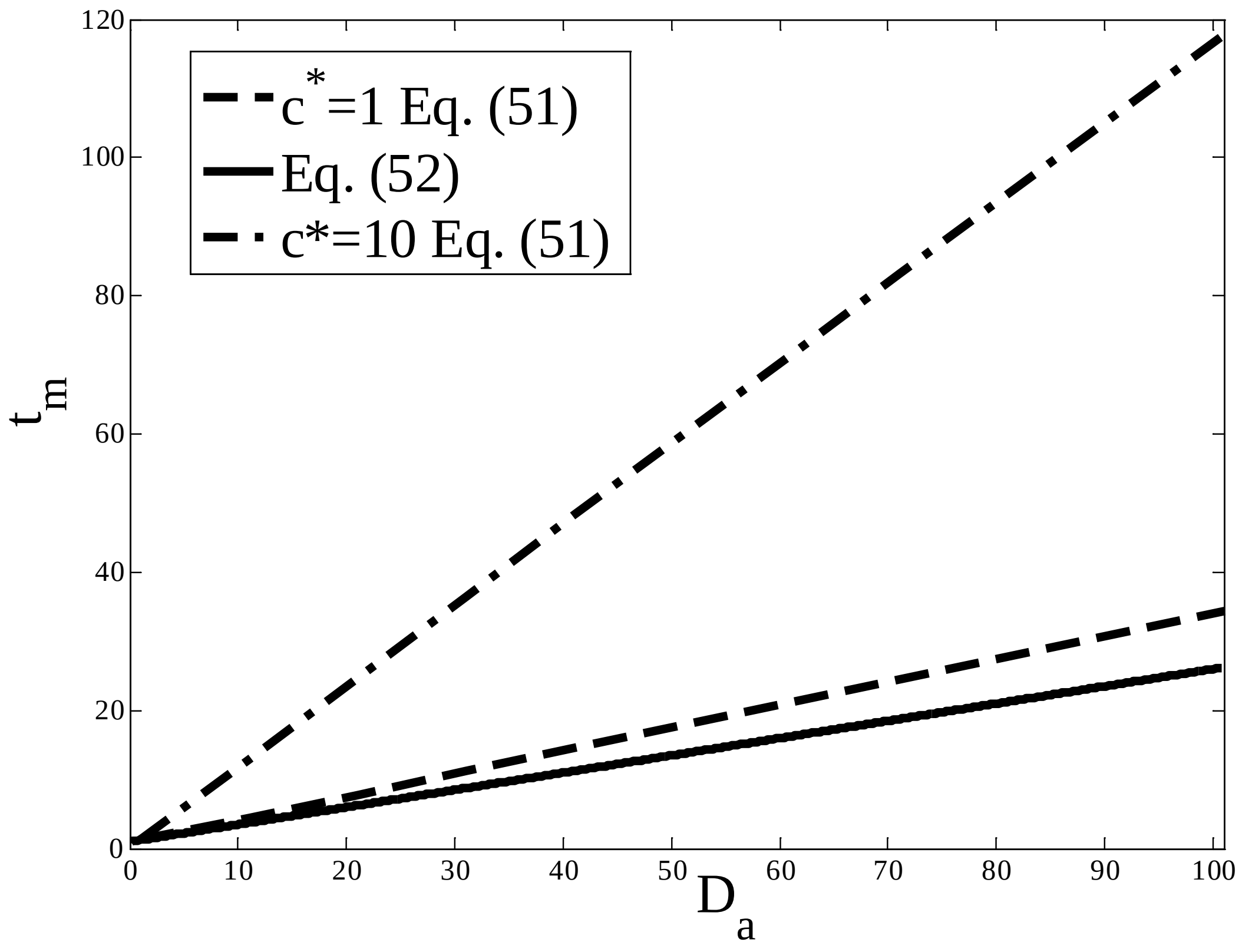